\documentstyle[twoside,11pt]{article}
\normalsize
\def\greaterthansquiggle{\raise.3ex\hbox{$>$\kern-.75em\lower1ex\hbox{$\sim$}}}
\def\lessthansquiggle{\raise.3ex\hbox{$<$\kern-.75em\lower1ex\hbox{$\sim$}}}
\newcommand{\bdi}{\begin{displaymath}}
\newcommand{\edi}{\end{displaymath}}
\newcommand{\bfi}{\begin{figure}}
\newcommand{\efi}{\end{figure}}

\newcommand{\beq}{\begin{equation}}
\newcommand{\eeq}{\end{equation}}

\newcommand{\gaM}{\gamma^{\mu}}

\newcommand{\beqa}{\begin{eqnarray}}
\newcommand{\eeqa}{\end{eqnarray}}

\newcommand{\ra}{\rightarrow}

\def\au{{\setbox0=\hbox{\lower1.36775ex%
\hbox{''}\kern-.05em}\dp0=.36775ex\hskip0pt\box0}}
\def\ao{{}\kern-.10em\hbox{``}}

\newcommand{\dsla}{\partial\hspace{-6pt} /  }

\newcommand{\ddsla}{\partial\hspace{-4.6pt} /  }

\newcommand{\AAsla}{A\hspace{-5pt}  /  }

\begin{document}
\bibliographystyle{plain}

\begin{titlepage}
\begin{flushright}
BUTP-95/27
\end{flushright}
\begin{center}

{\Large The Schwinger mass in the massive Schwinger model}\\

\bigskip

Christoph Adam \\
Institut f\"ur theoretische Physik, Universit\"at Bern \\
Sidlerstra\ss e 5, CH-3012 Bern, Switzerland$^*)$ \\

\bigskip

\today \\

\bigskip

{\bf Abstract} \\

\bigskip

 We derive a systematic procedure to compute Green functions for the
massive Schwinger model via a perturbation expansion in the fermion
mass. The known exact solution of the massless Schwinger model is
used as a starting point.

We compute the corrections to the Schwinger mass up to second order.

\vfill

\end{center}
$^*)${\footnotesize permanent address: Institut f\"ur theoretische Physik,
Universit\"at Wien \\
Boltzmanngasse 5, 1090 Wien, Austria \\
email address: adam@pap.univie.ac.at}
\end{titlepage}

\section{Introduction}

The massless Schwinger model is wellknown to be equivalent to a theory of a
free massive boson with Schwinger mass $\mu_0^2 = \frac{e^2}{\pi}$
(\cite{Sc1}). This massive field is formed via the chiral anomaly and may
be interpreted as a fermion--antifermion bound state (\cite{CKS}, \cite{IP1},
\cite{ABH}, \cite{Diss}, \cite{Le1}). Besides, the massless Schwinger model
shows other nontrivial features like fermion condensate, instantons and
nontrivial vacuum structure ($\theta$ vacuum) (\cite{LS1}--\cite{IP1},
\cite{DSEQ}--\cite{Diss}, \cite{Sm1}--\cite{KS1}, \cite{tH1}).
In the massless Schwinger model physical quantities do not
depend on the vacuum angle $\theta$.

The massive Schwinger model is different in some repects. First, it is no
longer exactly solvable (\cite{KS1}--\cite{Fry},
\cite{ABH}, \cite{Diss}). The physical state -- the massive boson -- is no
longer free, and its mass acquires corrections due to the interaction.
Instanton--like gauge fields and a nontrivial vacuum structure persist to be
present, and, in addition, physical quantities now depend on the vacuum angle
$\theta$. The fermion condensate, too, acquires corrections due to the fermion
mass (\cite{MSSM}).

Here we show how to compute Green functions for the massive Schwinger model
within the Euclidean path integral formalism, using a perturbation expansion
in the fermion mass. (The existence and finiteness of the mass perturbation
theory for the massive Schwinger model was proven in \cite{FS1}). For this
purpose we use the known exact solution of the massless Schwinger model as a
starting point. From the perturbation expansion we explicitly calculate
corrections to the Schwinger mass up to second orcer in the fermion mass $m$,
for a general vacuum angle $\theta$.

\section{Exact solution of the massless case}

The vacuum functional (and Green functions) of the massive Schwinger model may
be inferred from $n$--point functions of the massless Schwinger model via an
expansion in the fermion mass. Indeed, we may write for the Euclidean vacuum
functional ($k\ldots$ instanton number)
\beq
Z(m,\theta)=\sum_{k=-\infty}^{\infty}e^{ik\theta}Z_k (m)
\eeq
where
\bdi
Z_k (m)=N\int D\bar\Psi D\Psi DA^\mu_k e^{\int dx\Bigl[ \bar\Psi(i\ddsla
-e\AAsla_k +m)\Psi -\frac{1}{4}F_{\mu\nu}F^{\mu\nu}\Bigr] }
\edi
\bdi
=N\int D\bar\Psi D\Psi D\beta_k \sum_{n=0}^\infty \frac{m^n}{n!}\prod_{i=1}^n
\int dx_i \bar\Psi (x_i)\Psi (x_i)\cdot
\edi
\beq
\cdot \exp\{\int dx\Bigl[ \bar\Psi
(i\dsla -\epsilon_{\mu\nu}\gaM\partial^\nu \beta_k
)\Psi +\frac{1}{2}\beta_k \Box^2 \beta_k \Bigr] \}
\eeq
($A_\mu =\epsilon_{\mu\nu}\partial^\nu \beta$ corresponding to Lorentz gauge).
Therefore the perturbative computation of $Z(m,\theta)$ is traced back to the
computation of scalar VEVs ( $\langle\prod_i S(x_i )\rangle_0 $, $S(x)\equiv
\bar\Psi (x)\Psi (x)$ ) for the massless Schwinger model and some space time
integrations. It is useful to rewrite the scalar densities in terms of chiral
ones, $S(x)=S_+ (x)+S_- (x)$, $S_\pm \equiv \bar\Psi P_\pm \Psi$, because then
only a definite instanton sector $k=n_+ -n_-$ contributes to the VEV
$\langle\prod_{i=1}^{n_+} S_+ (x_i)\prod_{j=1}^{n_-} S_- (x_j)\rangle_0$. A
general VEV may be computed exactly (see e.g.
\cite{DSEQ}--\cite{Diss}, \cite{Zah}, \cite{MSSM}),
 \beq
\langle S_{H_1}(x_1)\cdots S_{H_n}(x_n)\rangle_0
=\Bigl( \frac{\Sigma}{2}\Bigr)^n \exp
\Bigl[ \sum_{i<j}(-)^{\sigma_i \sigma_j}4\pi D_{\mu_0} (x_i -x_j)\Bigr]
\eeq
where $\sigma_i =\pm 1$ for $H_i =\pm$, $D_{\mu_0}$ is the massive scalar
propagator,
\beq
D_{\mu_0}(x)=-\frac{1}{2\pi}K_0 (\mu_0 |x|), \quad \tilde
D_{\mu_0}(p)= \frac{-1}{p^2 +\mu_0^2},
\eeq
($K_0\ldots$ McDonald function) and $\Sigma$ is the fermion condensate
of the massless Schwinger model,
\beq
\Sigma =\langle\bar\Psi \Psi\rangle_0 =\frac{e^\gamma}{2\pi}\mu_0
\eeq
($\gamma\ldots$ Euler constant). The index 0 for $\mu_0$ indicates that it is
the order zero result, the index 0 for the VEVs means that they are with
respect to the massless Schwinger model. From this $Z(m,\theta)$ may be
computed (see \cite{MSSM} for details,
\cite{LSm} for its physical implications),
\bdi
Z(m,\theta)= e^{V\alpha (m,\theta)},
\edi
\beq
\alpha (m,\theta)=m\frac{\Sigma}{2}2\cos\theta + m^2
\Bigl(\frac{\Sigma}{2}\Bigr)^2 (E\cos 2\theta +F)+o(m^3 )
\eeq
($V\ldots$ space time volume) where
\bdi
E=\int d^2 xE(x) \equiv \int d^2 x(e^{-2K_0 (\mu_0
|x|)}-1)=-8.9139\cdot\frac{1}{\mu_0^2}
\edi
\beq
F=\int d^2 xF(x) \equiv \int d^2 x(e^{2K_0 (\mu_0 |x|)}-1) = 9.7384\cdot
\frac{1}{\mu_0^2}
\eeq
and for $F$ a subtraction of a free field singularity is necessary.

In order to compute VEVs for the massive Schwinger model one has to insert the
corresponding operators into the path integral (1), (2) and divide by the
vacuum functional $Z(m,\theta)$:
\beq
\langle \hat O \rangle_m =\frac{1}{Z(m,\theta)} \langle \hat O
\sum_{n=0}^\infty \frac{m^n}{n!}\prod_{i=1}^n \int dx_i \bar\Psi (x_i) \Psi
(x_i) \rangle_0
\eeq
Via the normalization all volume factors cancel completely, as it certainly has
to be (we will explicitly see this in the computations). For the computation
of VEVs of scalar $(S)$ and chiral $(S_\pm)$ densities formula (3) suffices
and could be used e.g. for the computation of the fermion condensate $\langle
\bar\Psi\Psi\rangle_m$  (this however may be derived at once from (6), see
\cite{MSSM}).

For the computation of the Schwinger mass additional Green functions of the
massless Schwinger model are needed. It is wellknown that the vector current
correlator is the free massive propagator in the massless Schwinger model:
\beq
\langle J_\mu (x)J_\nu (y)\rangle_0 =\frac{1}{\pi}\epsilon_{\mu\mu'}
\partial_x^{\mu'} \epsilon_{\nu\nu'}\partial_y^{\nu'}D_{\mu_0}(x-y)
\eeq
Therefore, for a perturbative calculation we need the $n$--point functions
\bdi
\langle J_\mu (y_2)J_\nu (y_1)\prod_{i=1}^n S_{H_i}(x_i)\rangle_0 =
\frac{1}{\pi}\epsilon_{\mu\mu'}
\partial_x^{\mu'} \epsilon_{\nu\nu'}\partial_y^{\nu'}\cdot
\edi
\bdi
\cdot \Bigl(\frac{\Sigma}{2}\Bigr)^n \Bigl[D_{\mu_0}(y_1 -y_2) +4\pi
(\sum_{i=1}^n (-)^{\sigma_i} D_{\mu_0}(x_i -y_2))(\sum_{j=1}^n (-)^{\sigma_j}
D_{\mu_0}(x_j -y_1))\Bigr]\cdot
\edi
\beq
\cdot e^{4\pi \sum_{k<l}(-)^{\sigma_k \sigma_l}D_{\mu_0} (x_k -x_l)}
\eeq
which may be computed from the Euclidean path integral or from bosonization
(the latter method is easier).

\section{Computation of the Schwinger mass}

For the perturbative computation of the massive propagator we simply have to
insert successive orders of equ. (10) into (8). The factor $(\frac{1}{\pi}
\epsilon_{\mu\mu'} \partial_{y_2}^{\mu'}
\epsilon_{\nu\nu'}\partial_{y_1}^{\nu'} )$ is the same for all orders,
therefore we will ignore it in the sequel.

There are two terms from (10) to be inserted in first order, namely $S_H ,
H=\pm$, and four terms, $S_{H_1}S_{H_2}, H_i =\pm$, in second order. For the
moment we consider the case where all $H_i =+$; the other contributions can be
inferred from this one by a rearrangement of some signs. Up to second order,
we find ( $\langle J_\mu (y_2 )J_\nu (y_1)\rangle_m \equiv \frac{1}{\pi}
\epsilon_{\mu\mu'} \partial^{\mu'}_{y_2} \epsilon_{\nu\nu'}
\partial^{\nu'}_{y_1} T(y_1 ,y_2)$ )
\bdi
T(y_1 ,y_2)= \frac{1}{Z(m,\theta)}\Bigl[ D_{\mu_0}(y_1 -y_2) +
m\frac{\Sigma}{2}V D_{\mu_0}(y_1 -y_2) + 4\pi m\frac{\Sigma}{2}\int
dxD_{\mu_0} (x-y_2)D_{\mu_0}(x-y_1) +
\edi
\bdi
\frac{m^2}{2!}\Bigl( \frac{\Sigma}{2}\Bigr)^2 \int dx_1 dx_2 [D_{\mu_0}(y_1
-y_2)+ 4\pi (D_{\mu_0}(x_1 -y_2)+D_{\mu_0}(x_2 -y_2))\cdot
\edi
\beq
\cdot (D_{\mu_0}(x_1 -y_1) +
D_{\mu_0}(x_2 -y_1))]e^{4\pi D_{\mu_0}(x_1 -x_2)}\Bigr]
\eeq
Inserting equ. (6) for the vacuum functional $Z(m,\theta)$ and using the
perturbation formula of second order,
\bdi
\frac{a_0 +a_1 x+a_2 x^2 +o(x^3 )}{1+b_1 x+b_2 x^2 +o(x^3 )}=
\edi
\beq
a_0 +a_1 x+a_2 x^2 -a_0 b_1 x-a_0 b_2 x^2 -a_1 b_1 x^2 +a_0 b_1^2 x^2 +o(x^3 )
\eeq
we arrive at
\bdi
T(y_1 ,y_2)=D_{\mu_0}(y_1 -y_2)+4\pi m\frac{\Sigma}{2}\int dx D_{\mu_0}(x-y_2)
D_{\mu_0}(x-y_1) +
\edi
\bdi
\frac{m^2}{2!}\Bigl(\frac{\Sigma}{2}\Bigr)^2 4\pi \int dx_1 dx_2 [2D_{\mu_0}
(x_1 -y_2)D_{\mu_0}(x_1 -y_1)+2D_{\mu_0}(x_1 -y_1)D_{\mu_0}(x_2 -y_2)]\cdot
\edi
\bdi
\cdot (E(x_1 -x_2)+1) - 4\pi m^2 \Bigl(\frac{\Sigma}{2}\Bigr)^2 V\int dx
D_{\mu_0} (x-y_2)D_{\mu_0}(x-y_1)\edi
\bdi
=D_{\mu_0}(y_1 -y_2)+4\pi m\frac{\Sigma}{2}\int dx D_{\mu_0}(x-y_2)
D_{\mu_0}(x-y_1) +
\edi
\bdi
m^2\Bigl(\frac{\Sigma}{2}\Bigr)^2 4\pi E \int
dxD_{\mu_0}(x-y_2)D_{\mu_0}(x-y_1)+
\edi
\beq
m^2 \Bigl(\frac{\Sigma}{2}\Bigr)^2 4\pi \int dx_1 dx_2 D_{\mu_0}(x_1 -y_1)
D_{\mu_0}(x_2 -y_2)(E(x_1 -x_2)+1)
\eeq
where $E,E(x)$ are given in (7) and we used the $x\ra -x$ symmetry of all
occurring functions. Observe that, as claimed, all volume factors $V$ have
dropped out.

However, expression (13) is not yet the desired result for further
computation, there is still one unwanted term. The last line of expression (13)
consists of two terms, due to the factor $(E(x_1 -x_2)+1)$. The second one,
proportional to 1, reads
\beq
m^2 \Bigl( \frac{\Sigma}{2}\Bigr)^2 4\pi\int dx_1 D_{\mu_0}(x_1 -y_1)\int dx_2
D_{\mu_0} (x_2 -y_2)
\eeq
and obviously does not contribute to the $y_1 -y_2$ correlation. It is a
disconnected part stemming from the VEVs
\beq
\langle J_\mu (y_i)S_+ (x_j)\rangle_0 \sim
\epsilon_{\mu\mu'}\partial_{y_i}^{\mu'} D_{\mu_0}(x_j -y_i)
\eeq
and must be subtracted from (13).

Looking for the moment at the first order correction only, it is very easy to
find the Schwinger mass correction:
\bdi
\Box_{y_1}T^{(1)}(y_1 ,y_2)=\delta (y_1 -y_2)+\mu_0^2 D_{\mu_0}(y_1 -y_2)+
\edi
\bdi
4\pi m\frac{\Sigma}{2}(D_{\mu_0}(y_1 -y_2)+\mu_0^2 \int dxD_{\mu_0}(x-y_2)
D_{\mu_0}(x-y_1))
\edi
\beq
\equiv \delta (y_1 -y_2)+\mu_1^2 D_{\mu_1}(y_1 -y_2)+o(m^2 )
\eeq
where
\beq
\mu_1^2 =\mu_0^2 +4\pi m\frac{\Sigma}{2}
\eeq
To obtain the second order result, we rewrite expression (13), without the
disconnected part, in momentum space and substitute all functions by their
Fourier transforms (thereby the convolutions turn into products):
\bdi
\tilde T(p)=\frac{-1}{p^2 +\mu_0^2}+4\pi m\frac{\Sigma}{2}\frac{1}{(p^2
+\mu_0^2)^2} +4\pi m^2 \Bigl(\frac{\Sigma}{2}\Bigr)^2
\frac{1}{(p^2 +\mu_0^2)^2} (E+\tilde E(p))=
\edi
\bdi
\frac{-1}{p^2 +\mu_0^2}(1-4\pi m\frac{\Sigma}{2}\frac{1}{p^2 +\mu_0^2}- 4\pi
m^2 \Bigl(\frac{\Sigma}{2}\Bigr)^2 (E+\tilde E(p))\frac{1}{p^2 +\mu_0^2})=\edi
\beq
\frac{-1}{p^2 +\mu_0^2 +4\pi m\frac{\Sigma}{2} +4\pi m^2 (\frac{\Sigma}{2})^2
(E+\tilde E(p))+(4\pi m \frac{\Sigma}{2})^2 \frac{1}{p^2 +\mu_0^2}}
+o(m^3)
\eeq
Therefore, for finding the mass pole, $p^2$ has to obey the self consistency
equation (after a rescaling ${p'}^2 =\frac{p^2}{\mu_0^2}$, $E' =E(\mu_0^2
\equiv 1)=\mu_0^2 E$ etc.)
\beq
{p'}^2 =-1-4\pi\frac{m}{\mu_0}\frac{\Sigma}{2\mu_0}-4\pi \frac{m^2}{\mu_0^2}
\Bigl( \frac{\Sigma}{2\mu_0}\Bigr)^2 [E' +\tilde E'(p') +\frac{4\pi}{{p'}^2
+1}]
\eeq
The second order part (the term in square brackets) may be rewritten like
\bdi
[\cdots ]=\int d^2 x[e^{-2K_0 (|x|)}-1+e^{ip' x}(e^{-2K_0 (|x|)}-1+ 2K_0
(|x|)]=
\edi
\bdi
\int_0^\infty dr r[2\pi (e^{-2K_0 (r)}-1) + \int_0^{2\pi} d\theta
e^{i|p'|r\cos\theta} (e^{-2K_0 (r)}-1+2K_0 (r))]=
\edi
\beq
2\pi \int_0^\infty dr r[e^{-2K_0 (r)}-1 + J_0 (|p'|r)(e^{-2K_0 (r)}-1+ 2K_0
(r))]
\eeq
where $J_0$ is the Bessel function of the first kind. This expression behaves
well around $|p'|=i$ and therefore we may set $|p'|=i$ because deviations
from this value are of higher order in $m$. Using $I_0 (r)=J_0 (ir)$ we find
\beq
{p'}^2=-1-4\pi\frac{m}{\mu_0}\frac{\Sigma}{2\mu_0} -8\pi^2 \frac{m^2}{\mu_0^2}
\Bigl(\frac{\Sigma}{2\mu_0} \Bigr)^2 \cdot A
\eeq
\bdi
A:= \int_0^\infty dr r[e^{-2K_0 (r)}-1 + I_0 (r)(e^{-2K_0 (r)}-1+2K_0 (r))]
\edi
\beq
A=-0.6599
\eeq
This result was computed for the positive chirality densities $S_+$. For the
pure negative chirality densities $S_-$ the result is completely identical.
However in second order there are mixed products $S_+ S_- ,S_- S_+$, too. Both
of them lead to the same result, where the above integral $A$ is substituted
by a similar expression $B$,
\bdi
B:=\int_0^\infty drr[e^{+2K_0 (r)}-1 + I_0 (r)(-e^{+2K_0 (r)}+1+2K_0 (r))]
\edi
\beq
B=1.7277
\eeq
In this expression the nice feature of cancellation of UV divergencies occurs.
Indeed, both $e^{2K_0 (r)}$ and $-I_0 (r)e^{2K_0 (r)}$ diverge like
$\frac{1}{r^2}$ for small $r$ (this divergency corresponds to the free fermion
field divergency of the underlying theory), but obviously the divergencies
cancel each other. In fact, this cancellation was already observed twenty
years ago in \cite{KS1} within a bosonization approach, and on general grounds
it should continue to hold for higher orders.

Collecting all results and multiplying each contribution with its
corresponding $\theta$ factor
($S_\pm \ra e^{\pm i\theta}$, see (1--3)), we find
for the Schwinger mass in second order
\beq
-{p'}^2 \equiv \frac{\mu_2^2}{\mu_0^2}=1+8\pi
\frac{m}{\mu_0}\frac{\Sigma}{2\mu_0} \cos\theta + 16\pi^2 \frac{m^2}{\mu_0^2}
\Bigl( \frac{\Sigma}{2\mu_0}\Bigr)^2 (A\cos 2\theta + B)
\eeq
or, inserting all numbers (remember
$\frac{\Sigma}{2\mu_0}=\frac{e^\gamma}{4\pi}$, equ. (5))
\beq
\mu_2^2 =\mu_0^2 (1 + 3.5621\cdot \frac{m}{\mu_0}\cos\theta + 5.4807 \cdot
\frac{m^2}{\mu_0^2} - 2.0933\cdot\frac{m^2}{\mu_0^2}\cos 2\theta )
\eeq
which is our final result.

For the special case $\theta =0$ our result (24) precisely coincides with
the result in \cite{Vary}, where the second order correction for $\theta =0$
was computed within bosonization and using near light cone coordinates. In the
same article this result was compared with a lattice calculation
(\cite{Crew}), and a good agreement is obtained within the range
of the expansion
parameter $\frac{m}{\mu_0}$ where the lattice calculations were performed.

\section{Summary}

We have demonstrated a general method of computing $n$--point functions for
the massive Schwinger model in mass perturbation theory within the Euclidean
path integral formalism, using the known exact solution of the massless
Schwinger model as a starting point.

All $n$--point functions exist
perturbatively and are finite in the infinite volume limit.

Using this approach we were able to compute the mass
perturbation corrections to
the Schwinger mass up to second order, and almost all calculations could be
done analytically. This feature does not persist to hold for higher orders.
Already in third order an equation analogous to (19) will be a true self
consistency equation that can be solved only numerically.

For $\theta =0$, the result (24) is well reproduced by lattice calculations
(\cite{Vary}).

The numerical calculations in this article were done with Mathematica 2.2.

\section*{Acknowledgements}
The author thanks H. Leutwyler for very helpful discussions and the members of
the Institute of Theoretical Physics at Bern University, where this work was
done, for their hospitality.

This work was supported by a research stipendium of the University of Vienna.


\begin{thebibliography}{9999}
\bibitem{Sc1}
J. Schwinger, Phys. Rev. {\em 128} (1962) 2425
\bibitem{LS1}
J. Lowenstein, J. Swieca, Ann. Phys. {\em 68} (1971) 172
\bibitem{Jay}
C. Jayewardena, Helv. Phys. Acta {\em 61} (1988) 636
\bibitem{SW1}
I. Sachs, A. Wipf, Helv. Phys. Acta {\em 65} (1992) 653
\bibitem{IP1}
N. P. Ilieva, V. N. Pervushin, Sov. J. Part. Nucl. {\em 22} (1991) 275
\bibitem{CKS}
A. Casher, J. Kogut, P. Susskind, Phys. Rev. {\em D10} (1974) 732
\bibitem{DSEQ}
C. Adam, preprint UWThPh-1994-39; HEP-PH 9501273
\bibitem{Adam}
C. Adam, Z. Phys. {\em C63} (1994) 169
\bibitem{Diss}
C. Adam, thesis Universit\"at Wien 1993
\bibitem{ABH}
C. Adam, R. A. Bertlmann, P. Hofer, Riv. Nuovo Cim. {\em 16}, No
8 (1993)
\bibitem{Sm1}
A. V. Smilga, Phys. Rev. {\em D46} (1992) 5598
\bibitem{Sm2}
A. V. Smilga, Phys. Rev. {\em D49} (1994) 5480
\bibitem{KS1}
J. Kogut, P. Susskind, Phys. Rev. {\em D11} (1975) 3594
\bibitem{CJS}
S. Coleman, R. Jackiw, L. Susskind, Ann. Phys. {\em 93} (1975) 267
\bibitem{Co1}
S. Coleman, Ann. Phys. {\em 101} (1976) 239
\bibitem{Fry}
M. P. Fry, Phys. Rev. {\em D47} (1993) 2629
\bibitem{FS1}
J. Fr\"ohlich, E. Seiler, Helv. Phys. Acta {\em 49} (1976) 889
\bibitem{GS1}
R. E. Gamboa Saravi, M. A. Muschietti, F. A. Schaposnik, J. E. Solomin,
Ann. Phys. {\em 157} (1984) 360
\bibitem{Le1}
H. Leutwyler, Helv. Phys. Acta {\em 59} (1986) 201
\bibitem{DR1}
W. Dittrich, M. Reuter, "Selected Topics ...", Lecture Notes in
Physics {\em Vol.244}, Springer, Berlin 1986
\bibitem{MSSM}
C. Adam, preprint BUTP-95/26, HEP-PH 9507279
\bibitem{Zah}
J. Steele, A. Subramanian, I. Zahed, HEP-TH 9503220
\bibitem{tH1}
G. t'Hooft, Phys. Rev. {\em D14} (1976) 3432, Phys. Rev. Lett. {\em 37} (1976)
8
\bibitem{LSm}
H. Leutwyler, A. V. Smilga, Phys. Rev. {\em D46} (1992) 5607
\bibitem{Vary}
J. P. Vary, T. J. Fields, H. J. Pirner, preprint ISU-NP-94-14, HEP-PH 9411263
\bibitem{Crew}
D. P. Crewther, C. J. Hamer, Nucl. Phys {\em B170} (1980) 353

\end{thebibliography}
\end{document}